\documentclass[journal=nalefd,manuscript=letter]{achemso}

\usepackage[version=3]{mhchem} 

\usepackage{nohyperref}  
\usepackage{url} 
\author{Giuseppe Romano}
\affiliation[MIT, Department of Mechanical Engineering]{Department of Mechanical Engineering, Massachusetts Institute of Technology, 77 Massachusetts Avenue, Cambridge (MA), 02139}
\alsoaffiliation[Boston College]{Department of Physics, Boston College, Chestnut Hill, MA 02467, USA}
\email{romanog@mit.edu}
\author{Keivan Esfarjani}
\affiliation[Rutgers University]{Department of Mechanical and Aerospace Engineering, Rutgers University, Piscataway, 08854 NJ, USA}
\author{David A. Strubbe}
\affiliation[MIT, Department of Materials Science and Engineering]{Department of Materials Science, Massachusetts Institute of Technology, 77 Massachusetts Avenue, Cambridge (MA), 02139}
\author{David Broido}
\affiliation[Boston College]{Department of Physics, Boston College, Chestnut Hill, MA 02467, USA}
\author{Alexie M. Kolpak}
\affiliation[MIT, Department of Mechanical Engineering]{Department of Mechanical Engineering, Massachusetts Institute of Technology, 77 Massachusetts Avenue, Cambridge (MA), 02139}

\title{Temperature-dependent thermal conductivity in nanoporous materials
studied by the Boltzmann Transport Equation}

\begin{document}


\begin{abstract}

Nanostructured materials exhibit low thermal conductivity because of the additional scattering due to
phonon-boundary interactions. As these interactions are highly sensitive to the mean free path (MFP) of a given phonon mode, MFP distributions in nanostructures can be dramatically distorted relative to bulk. Here we calculate the MFP distribution in periodic nanoporous Si for different temperatures, using the recently developed MFP-dependent Boltzmann Transport Equation.
After analyzing the relative contribution of each phonon branch to
thermal transport in nanoporous Si, we find that at room temperature optical phonons
contribute $18\%$ to heat transport, compared to $5\%$ in bulk Si. Interestingly, we observe a steady thermal conductivity in the nanoporous materials over a temperature range $200~\mathrm{K}~<~T~<~300~\mathrm{K}$, which we attribute to the ballistic transport of acoustic phonons with long intrinsic MFP. These results, which are also consistent with a recent experimental study, shed light on the origin of the reduction of thermal
conductivity in nanostructured materials, and could contribute to multiscale
heat transport engineering, in which the bulk material and geometry are
optimized concurrently.

\end{abstract}

\section{Introduction}
The quest for high-efficiency thermoelectric materials may be advanced by using the ability of nanostructures to suppress heat transport by several orders of magnitude with respect to bulk without degrading electrical transport significantly~\cite{20661949}. This phenomenon is based on the fact that phonon mean free paths (MFPs) are generally larger than electron MFPs; consequently, heat transport exhibits stronger size effects. The extent of the suppression of phonon transport depends on the ratio between the intrinsic phonon MFP and the characteristic length of the nanostructure, $L_c$. This ratio is known as the Knudsen number ($Kn$). When $L_{\mathrm c}$ is much smaller than MFP, \textit{i.e.} for small $Kn$, phonon interactions with boundaries are negligible. In this regime, heat transport reduction is only due to geometrical effects, such as material removal in nanoporous materials, while phonon-boundary scattering is minimal. Therefore, heat transport is dominated by intrinsic scattering. On the other hand, for high $Kn$ scattering is dominated by phonon-boundary interactions. Within this regime, the phonon MFPs in the nanostructure approach $L_c$ and the phonons are considered to travel ballistically. The intermediate regime (\textit{i.e.} $Kn \approx 1$) is often refered to as the quasi-ballistic regime. 

This analysis implicitly assumes single-phonon mode materials, but in most materials, there is a wide distribution of phonon MFPs, which in some cases span several orders of magnitude. For example, first-principles calculations for Si show that about half of the heat is carried by phonons with MFPs larger than $1~\mathrm{\mu m}$~\cite{esfarjani2011heat}. Recent experimental measurements showing a reduction in thermal conductivity of Si membranes with microscale pores~\cite{song2004thermal} provide support for these computational results. Together, they suggest that an accurate analysis of thermal transport in nanostructures should include the actual bulk MFP distribution.

In bulk Si, the optical and acoustic phonons have very different MFP distributions. Optical phonons have relatively low MFPs because their dispersion curves are flatter than those of acoustic phonons, which by contrast have large MFPs. This effect has important consequences on thermal conductivity. First-principles calculations show that optical phonons contribute only $5\%$ to the total thermal conductivity of Si, while the $95\%$ is dominated by acoustic phonons~\cite{esfarjani2011heat}. As a result, optical phonons are often neglected when calculating nanoscale heat transport in Si. However, in nanostructures, heat carried by optical phonons is slightly lowered while acoustic phonons can be strongly suppressed, making the two contributions comparable. 

A simple model for phonon-boundary scattering that neglects intrinsic scattering and assumes that phonons travel ballistically was devised by Casimir in 1938~\cite{casimir1938note}. Within the Casimir approach, the MFPs are considered to be the same as the characteristic length. By using the Casimir model, Tian \textit{et al}~\cite{Tian_2011} concluded that optical phonons in Si nanowires contribute over $20\%$ to the total thermal conductivity at room temperature. When dealing with complex boundaries, the Casimir approach fails for two reasons~\cite{marconnet2013casimir}. First, it assumes that the characteristic length is known a priori, while in most materials with complex geometry this quantity is unknown. Secondly, a portion of the MFP distribution may lie in the diffusive or quasi-ballistic regime.  

In this work, we use the MFP-dependent Boltzmann Transport Equation (MFP-BTE)~\cite{Romano_2015} to calculate heat transport in nanoporous materials and provide the relative contribution of each phonon branch to the thermal thermal conductivity as a function of temperature. The use of the BTE enables treatment of complex geometries with a good level of predictive power. We focus on nanoporous Si (np-Si) with aligned pores with square cross section and a periodicity of $10~\mathrm{nm}$, and we consider the temperature range $100~\mathrm{K}-300~\mathrm{K}$. We show that at room temperature the thermal conductivity in np-Si is suppressed by more than one order of magnitude with respect to bulk Si, with longitudinal optical (LO) phonons contributing nearly $20\%$ to the total heat transport. This result is in agreement with the previous qualitative discussion. Further, we find that the thermal conductivity of np-Si exhibits a plateau over the temperature range $200~\mathrm{K}-300~\mathrm{K}$. We demonstrate that this arises from two effects: first, as most of the acoustic phonons travel ballistically because of their large $Kn$, their MFPs in np-Si are constrained by the characteristic length of the material. Second, in this temperature range, the heat carried by optical phonons changes weakly with temperature in bulk Si itself, inducing similar behavior to np-Si. By revealing the microscopic mechanisms leading to the reduction in heat conduction, our findings may enable new approaches for engineering high-efficiency thermoelectric devices.

\section{Results and discussion}

To compute the reduction of heat transport in nanostructures, we employ the concept of the ``suppression function,'' $S(\Lambda)$, which defines the departure from diffusive transport in terms of the MFP distribution function~\cite{Romano_2015},
\begin{equation}
S(\Lambda)=\frac{K^{\mathrm {nano}}_p(\Lambda)}{K^{\mathrm{bulk}}_p(\Lambda)},
\end{equation}
where $K_p^{\mathrm{bulk}}(\Lambda)$ and $K_p^{\mathrm{nano}}(\Lambda)$ are the bulk MFP distributions for branch $p$ in bulk Si and np-Si, respectively. Within the relaxation time approximation, the effective thermal conductivity for each phonon branch can be written as 
\begin{equation}\label{ptc}
 \kappa^{nano}_p = \int_0^{\infty} K^{nano}_p(\Lambda) d\Lambda = \int_0^{\infty} K^{bulk}_p(\Lambda) S(\Lambda) d\Lambda.
\end{equation}
The total thermal conductivity is then given by $\kappa^{\mathrm{nano}} = \sum_p \kappa_p^{\mathrm{nano}}$. In the case of purely diffusive transport, the suppression function is MFP-independent and ~\ref{ptc} leads to the diffusive thermal conductivity $\kappa^{\mathrm{nano}} = \kappa^{\mathrm{bulk}}g$, where $g$ is a function that depends only on the material geometry and $\kappa^{\mathrm{bulk}}$ is the bulk thermal conductivity. The bulk MFP distribution at different temperatures can be obtained either experimentally through MFP reconstruction techniques~\cite{minnich2011thermal} or computationally. In this work we adopt a first-principles approach based on Density Functional Theory (DFT) and the linearized BTE~\cite{broido2007intrinsic,esfarjani2011heat}. The bulk MFP distribution is computed via $K_p^{\mathrm{bulk}}(\Lambda)=\frac{\partial \alpha_p^{\mathrm{bulk}}(\Lambda)}{\partial \Lambda}$, where $\alpha_p^{\mathrm{bulk}}(\Lambda)$ is cumulative thermal conductivity. We recall that the cumulative thermal conductivity is the thermal conductivity of phonons whose MFPs are below a given $\Lambda$~\cite{1995,dames2006thermal}. The intrinsic MFP for a given phonon mode is defined as $\Lambda=|\mathbf{v}|\tau^{\mathrm{bulk}}$. We note that $\alpha_p^{\mathrm{bulk}}(\Lambda)$ does not include boundary scattering. For this reason, in rest of this study, we will refer to such a result as bulk-BTE.

The suppression function can be obtained in different ways, depending on the system and the required accuracy. In some cases, such in nanowires and thin films, $S(\Lambda)$ can be obtained analytically within a reasonable level of accuracy~\cite{Yang_2013}. However, most of the analytical derivations are based on the ``gray'' approximation, which assumes phonon dispersions described by a single group velocity. Furthermore, formulae for the suppression function are limited to simple geometries. In this work we therefore employ a recently developed formulation of the BTE that requires only the bulk MFP distribution $K^{\mathrm{bulk}}_p(\Lambda)$~\cite{Romano_2015}. This method, MFP-BTE, has the same accuracy as the the commonly used frequency dependent approach (FD-BTE), provided that we consider small applied temperature gradients, $\Delta T/L$, where $\Delta T$ is the applied difference of temperature and $L$ is the distance between the hot and cold contact. The key equation of the MFP-BTE consists of the integro-differential equation
\begin{equation}\label{mfp}
\Lambda \mathbf{s}\cdot\nabla \tilde{T}(\Lambda) + \tilde{T}(\Lambda)=\gamma\sum_p \int_0^{\infty}\frac{K^{\mathrm{bulk}}_p(\Lambda')}{\Lambda'^2}<\tilde{T}(\Lambda')>  d\Lambda',
\end{equation}
where $\tilde{T}(\Lambda)$ represents the normalized temperature associated with phonons having MFP $\Lambda$, given by $\tilde{T}(\mathbf{r},\Lambda)=\frac{T(\mathbf{r},\omega,p)-T_0}{\Delta T}$. In ~\ref{mfp}, $\mathbf{s}$ is the phonon propagation direction and $\gamma$ a material property given by $\gamma = \left[\sum_p \int_0^\infty  \frac{K_p^{\mathrm{bulk}}}{\Lambda^2} d\Lambda \right]^{-1} = 2.2739~\rm{x}~10^{-17}~\rm K W^{-1} m^{-3}$ for Si. The notation $<x>$ stands for an angular average. The right-hand-side of ~\ref{mfp} is equal to $\frac{T_L(\mathbf{r})-T_0}{\Delta T}$, where $T_L$ is the effective lattice temperature~\cite{Romano_2015}, which does not depends on $\Lambda$ explicitly, and provides an average of the local energy of phonons. Once ~\ref{mfp} is solved, the suppression function can be computed via
\begin{equation}
S(\Lambda) = \frac{3L}{\Lambda A}\int_{\Gamma} <\tilde{T}\mathbf{s}\cdot \mathbf{n}_s> dx^2,
\end{equation} \label{conn}
where $\Gamma$ is the surface of the hot contact having normal $\mathbf{n}_s$ and area $A$. 

Our simulation domain consists of a square unit cell, which we choose to have a size of $L~=~10~\mathrm{nm}$. The unit cell, which contains one square pore, is subjected to a difference of temperature $\Delta T~=~1~\mathrm{K}$ about $T_0$. Periodic boundary conditions are applied to both the longitudinal and transverse direction of heat flux, $\mathbf{n_f}$, which is enforced by applying a difference of temperature $\Delta T$ along $\mathbf{n_f}$, \textit{i. e.}
\begin{equation}\label{boundary}
\tilde{T}(\Lambda,\mathbf{s},\mathbf{r})-\tilde{T}(\Lambda,\mathbf{s},\mathbf{r}+\mathbf{P}) = (\mathbf{n}\cdot\mathbf{n_f})\Delta T,
\end{equation}
where $\mathbf{r}$ runs along the faces of the unit cell, $\mathbf{n}$ is the normal to the boundary pointing outside the domain and $\mathbf{P}$ is the periodicity vector. It is straightforward to show that along the direction perpendicular to the heat flux, no difference of temperature is imposed. We assume that we have an infinite material along the directions orthogonal to the pore plane. Moreover, we assume that phonons are scattered diffusively when interacting with the pores, \textit{i.e.}, the distribution of phonons moving away from a pore's surface is uniformly distributed in angular space. We note that this assumption must be used with caution, especially at very low temperatures, as some phonons can be reflected specularly depending on the roughness of the boundary. However, in this work we assume the surfaces have significant roughness such that specularity effects can be neglected in the considered temperature range. Details on the surface specularity effects on thermal transport can be found in Ref.~\cite{Romano_2012}.

\begin{figure}[h!]
\begin{center}
\includegraphics[width=0.7\columnwidth]{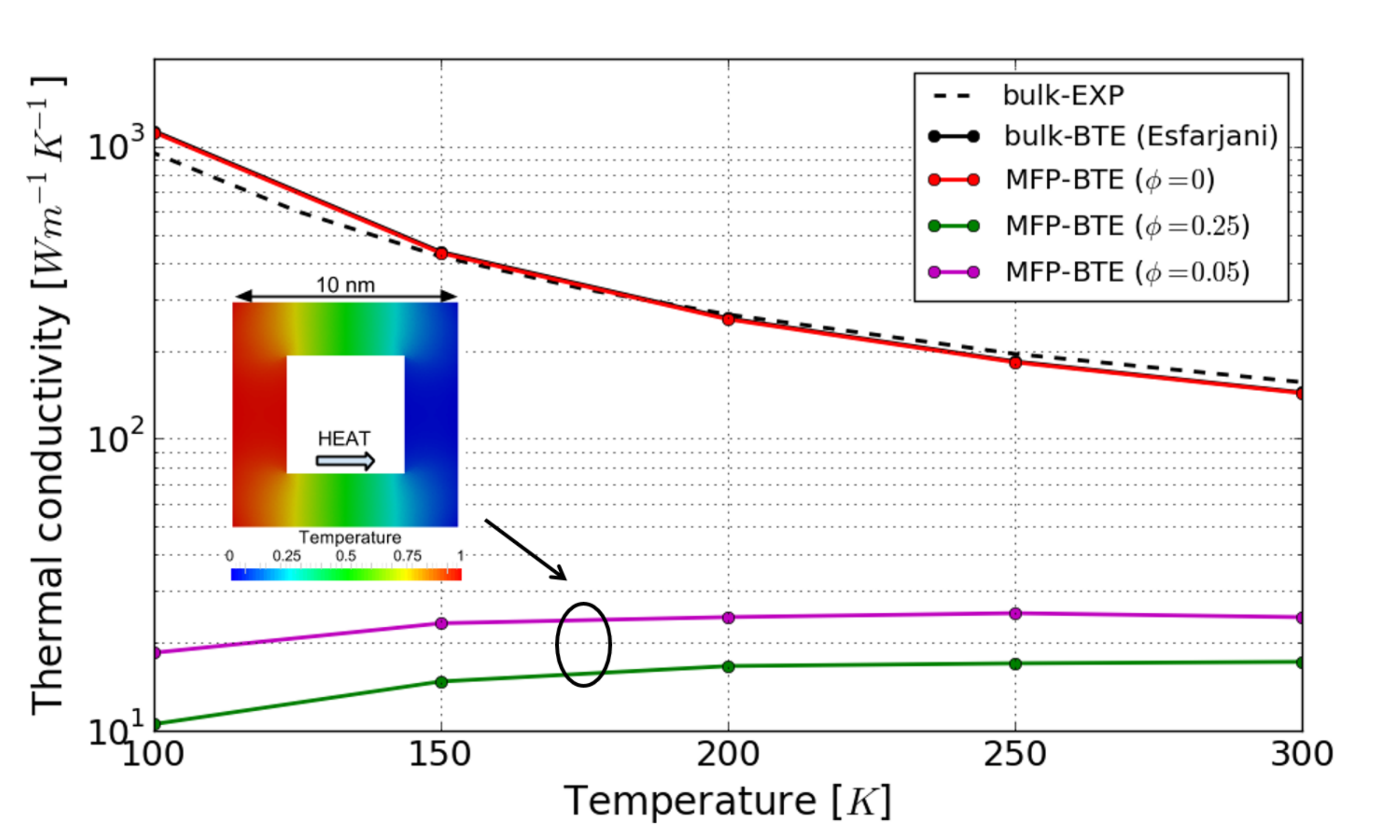}
\caption{The thermal conductivity for different temperatures of bulk Si and np-Si. For bulk Si, the experimental data~\cite{glassbrenner1964thermal}, bulk-BTE and MFP-BTE calculations are shown and agree with each other. For np-Si, we show the cases of porosity $\phi=0.05$ and $\phi=0.25$, computed by the MFP-BTE. In the inset, we show the normalized temperature map of the np-Si with porosity $\phi=0.25$.}\label{fig:figure1}
\end{center}
\end{figure}

We first validate the code with the case of zero porosity, \textit{i.e.}, bulk Si. Figure~\ref{fig:figure1} illustrates the good agreement between experimental data~\cite{glassbrenner1964thermal}, bulk BTE predictions, and MFP-BTE calculations of the thermal conductivity of bulk Si. The bulk case recovers purely diffusive transport, as there are no boundaries suppressing phonons. 

\begin{figure}[h!]
\begin{center}
\includegraphics[width=1.0\columnwidth]{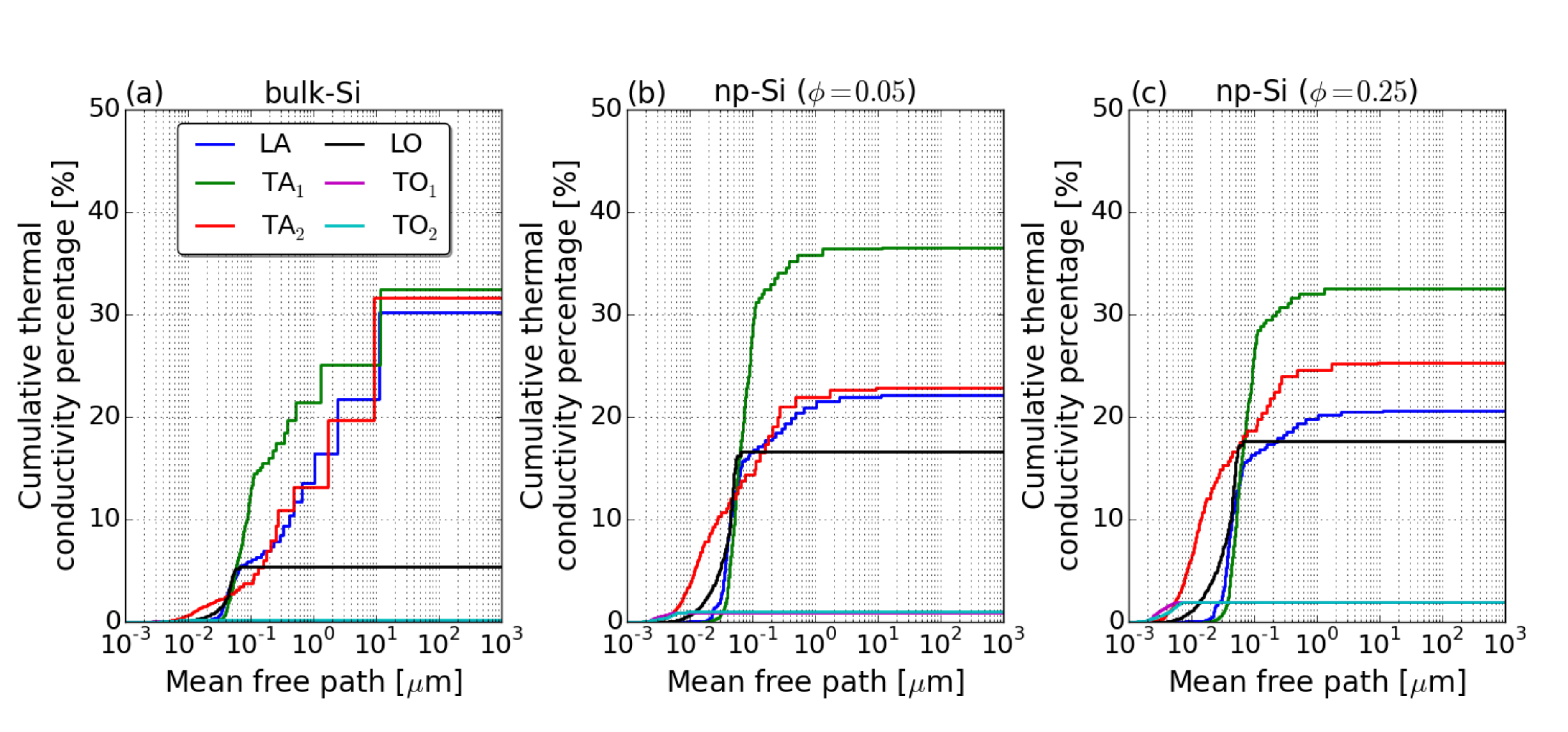}
\caption{Cumulative thermal conductivity at $T=300K$ for a) bulk Si, b) np-Si with porosity $\phi = 0.25$ and c) np-Si with porosity $\phi = 0.05$. All the values are normalized to the total thermal conductivity. The largest MFP contributing to heat transport for a given branch can been seen from the point where the relative cumulative thermal conductivity becomes flat.}\label{fig:figure2}
\end{center}
\end{figure}

Figure~\ref{fig:figure1} also shows the thermal conductivity for np-Si with porosities $\phi=0.05$ and $\phi=0.25$. At $T~=~300~\mathrm{K}$, we observe a reduction of more than one order of magnitude with respect to the bulk, in accordance with our previous studies~\cite{Romano_2014,Romano_2015}.   
We now analyze the relative contribution of each phonon branch to the total thermal conductivity. In Fig.~\ref{fig:figure2}-a, the normalized cumulative thermal conductivity for bulk Si at $T=300~\rm K$, computed by the the bulk BTE, is shown. As expected, acoustic phonons contribute most to the thermal conductivity, while optical phonon contributions are small. In particular, the two transverse acoustic (TA) branches and longitudinal acoustic (LA) branch contribute approximately one third each to the total thermal conductivity. The figure also shows that the longitudinal optical (LO) phonons contribute $5\% $ to thermal transport, while the transverse optical (TO) phonons have a negligible contribution. However, the LO phonons start to contribute significantly in np-Si, reaching $16\%$ of the total thermal conductivity in the case of $\phi=~0.05$ (Fig.~\ref{fig:figure2}-b). The TO contribution remains negligible.     

The roughly four-fold increase in the relative contribution of LO phonons can be better understood by analyzing the MFP distributions in relation to $L_c$. According to Ref.~\cite{Romano_2014}, in porous materials $L_c$ can be defined as the pore-pore distance in the direction orthogonal to thermal flux. The pore-pore distance in an array of square aligned pores is related to the porosity via $L_c =L( 1 - \sqrt{\phi})$, which leads to the values $5~\mathrm{nm}$ and $7.76~\mathrm{nm}$ for $\phi=0.25$ and $\phi=0.05$, respectively. The characteristic length dictates the transport regime of phonons with a given MFP. Figure~\ref{fig:figure2}-a shows that the maximum MFP of LO phonons contributing to the thermal conductivity is around $20~nm$, while acoustic phonons have MFPs up to $10~\rm \mu m$. As a result, optical phonons, which generally have MFPs similar to $L_c$, are less suppressed than acoustic phonons. For $\phi=0.25$, the characteristic length is even smaller and, consequently, the relative LO phonon contribution increases (up to $~18 \%$, as shown in Fig.~\ref{fig:figure2}-c). 

When $L_c$ is larger (\textit{e.g.}, $~100~\mathrm{nm}$), the effect of the nanostructure on optical phonons becomes negligible, but most acoustic phonons are still suppressed. In this case, it is possible to have a ``reversal effect,'' in which optical phonons are the main contribution to the thermal conductivity. For macroscopic samples, e.g. $L_c>~100~\rm \mu m$, the thermal conductivity approaches the value predicted by the Fourier model and the MFP distributions are restored to the bulk ones times the geometric factor, $g$, that depends only on the geometry. For aligned porous materials, the geometry factor can be well approximated by $g=\frac{1-\phi}{1+\phi}$~\cite{nan1997effective}. This approximation was validated against finite-element modeling of diffusive heat conduction~\cite{Romano_2012}. 

This finding has important consequences on the guidelines for optimizing nanostructured thermoelectric materials. Typically, the geometry of the nanostructure and the bulk thermoelectric materials are optimized separately. Here we suggest that both macro and nanoscale have to be considered concurrently. The following example helps clarify this point. Let us assume that we have two ``gray'' materials, $A$ and $B$, with average MFPs $\Lambda_A$ and $\Lambda_B$, respectively. We further assume that the thermal conductivity of material $B$ is larger than that of material $A$.  We consider a nanostructure with $\Lambda_A<<L_c<<\Lambda_B$. Material $B$ will undergo strong phonon suppression whereas heat transport in material $A$ will still be in the diffusive regime. It is clear therefore that, with a sufficiently large $\Lambda_B$, material $B$ exhibits lower thermal conductivity than that of material $A$, resulting more appealing for thermoelectrics. Similar conclusions can be drawn for non-gray materials.

\begin{figure}[h!]
\begin{center}
\includegraphics[width=0.7\columnwidth]{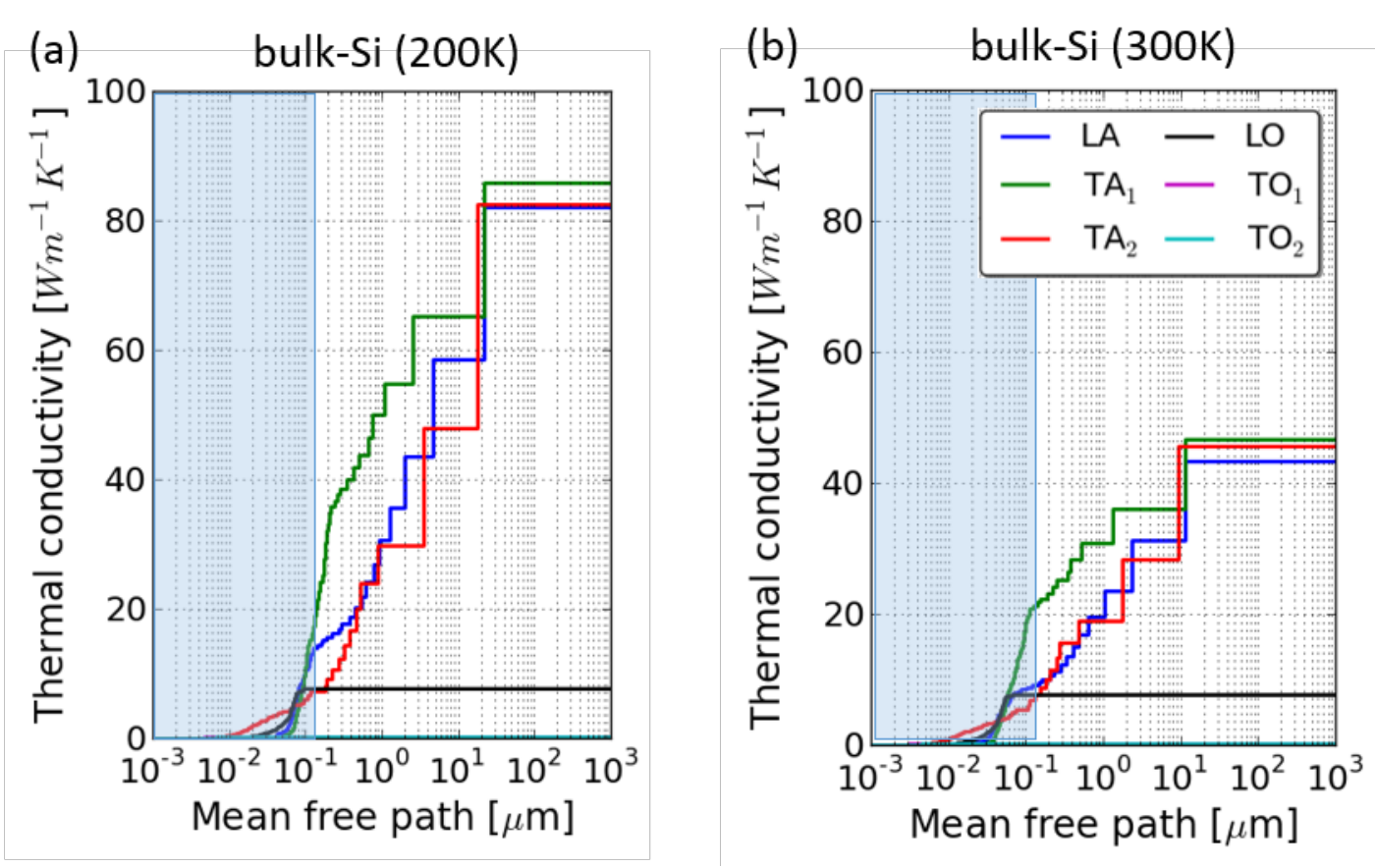}
\caption{MFP distributions of bulk Si at a) $T=200~K$ and b) $T=300~K$. In both panels, the shaded area shows the region of MFPs that are close to the characteristic size of the nanostructure, $L_c = 5~nm$. At low temperatures, the contribution to the thermal conductivity from long-MFP acoustic phonons rises. However, the ballistic regime constrains these MFPs to be equal to $L_c$.  
}\label{fig:figure3}
\end{center}
\end{figure}

\begin{figure}[h!]
\begin{center}
\includegraphics[width=0.7\columnwidth]{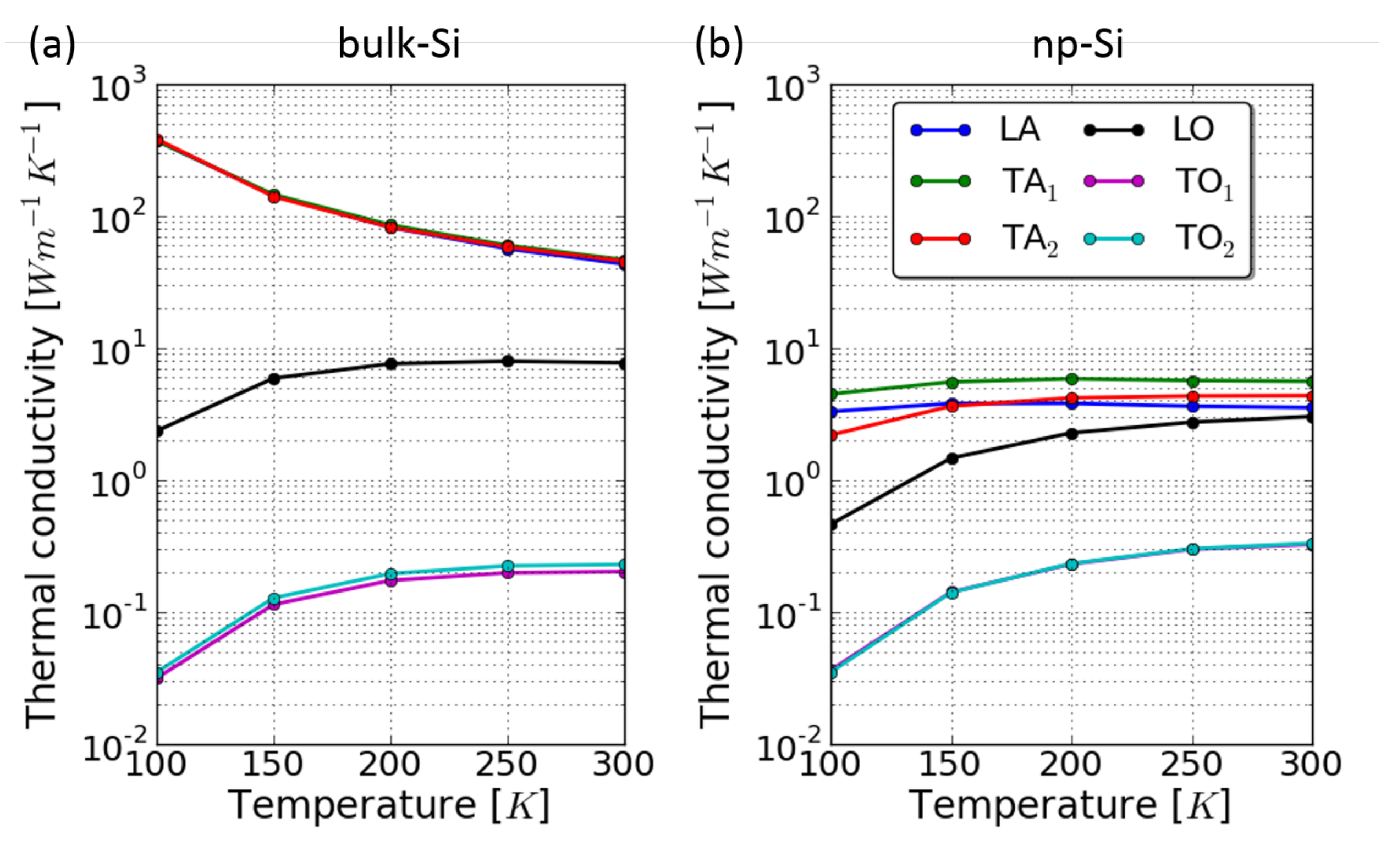}
\caption{Contribution of each phonon branch to the total thermal conductivity for a) bulk Si and b) np-Si. In the range $200~\mathrm{K}<T<300~\mathrm{K}$, the contributions to the thermal conductivity from each phonon branch in np-Si do not change significantly with temperature.
}\label{fig:figure4}
\end{center}
\end{figure}

We now investigate the temperature dependence of thermal conductivity in the range $200~\rm{K}~<~T~<~300~\rm{K}$.  All the results shown below refer to the case with $\phi=0.25$. Similar conclusions can be drawn for the case with $\phi=0.05$. As shown in Fig.~\ref{fig:figure1}, the thermal conductivity of np-Si exhibits little change in this temperature range, whereas it decreases as $1/T$ due to Umklapp scattering in bulk Si~\cite{broido2007intrinsic,esfarjani2011heat}. This behavior arises from the very large $Kn$ of acoustic phonons, which therefore travel ballistically. Within this regime, the phonon suppression function goes as $S(\Lambda)\approx L_c/\Lambda$~\cite{chen2005nanoscale}, and the contribution of acoustic phonons in np-Si becomes independent of the bulk MFP. According to Fig.~\ref{fig:figure3}, for temperatures as low as $200~\rm K$, the contribution to the total thermal conductivity from large MFP acoustic phonons becomes even greater.
Optical phonons, on the other hand, have MFPs close to $L_c$, and, in principle, their temperature dependence in bulk Si would affect their MFPs in np-Si. However, according to Fig.~\ref{fig:figure3}-a, for temperatures $200~K~<T<300~K$, heat carried by LO phonons in bulk Si does not change significantly with temperature because the increase in heat capacity is compensated by the decrease in scattering time~\cite{esfarjani2011heat}. Consequently, heat carried by LO phonons in np-Si does not change with temperature. Heat carried by TO phonons is negligible in both bulk Si and np-Si. These combined effects lead to the observed insensitivity of thermal conductivity to temperature in np-Si. Finally, as the temperature decreases in the range $100~K~<T<200~K$ because of the decrease of the heat capacity, heat carried by LO phonons in bulk Si starts to decrease, as shown in Fig.~\ref{fig:figure4}-a. As a consequence, their relative contributions to the thermal conductivity in np-Si decreases, as well.   

Our prediction of the temperature-dependence of thermal conductivity has been confirmed experimentally in a recent study, where heat transport has been measured in holey Si with periodicity of about $20~\mathrm{nm}$~\cite{lee2015ballistic}. We point out that our model, being based on BTE, calculates phonon dynamics within the particle picture, while the wave picture is only retained in the calculations of the bulk dispersion curves. As a consequence, our study suggests that in such systems incoherent scattering dominates thermal transport. The importance of coherence effects has been assessed in another recent study, where Monte Carlo simulations were used to compute thermal transport in np-Si membranes~\cite{ravichandran2014coherent}. Their conclusion was, however, that incoherent effects can fully explain the remarkably low thermal conductivity in silicon nanomeshes~\cite{yu_2010}. 


\section{Conclusions}

Using the MFP-BTE, we calculate the temperature dependence of thermal conductivity in np-Si. We quantify the contribution of optical phonons to thermal conductivity in np-Si with periodicity $10~\mathrm{nm}$, which at room temperature amounts to $~18 \%$. We also predict constant thermal conductivity over the range $200~\rm K<T<300~\rm K$, which has been recently observed experimentally. Our findings help further the understanding and manipulation of heat transport at the nanoscale for low thermal-conductivity applications such as thermoelectrics. We have also showed that the effectiveness of nanostructuring in reducing thermal transport does not depend on the bulk thermal conductivity but rather on the bulk MFP distributions of phonon branches. Consequently, our approach suggests that the search for better nanostructured thermoelectric materials has to involve the shape of the bulk cumulative thermal conductivity in relation with the material's geometry. In other words, the material optimization has to be done at both macro and nanoscale concurrently. 

\section{Methods}
We define the cumulative thermal conductivity in the bulk material as
\begin{equation}\label{cumulative}
\alpha^p_{bulk}(\Lambda) = \frac{1}{(2 \pi)^3}\int_{B.Z.}C_p(\mathbf{q}) v^2_{p,x}(\mathbf{q})\tau^{bulk}_p(\mathbf{q}) \, \Theta(\Lambda-\tau^{bulk}_p(\mathbf{q})|\mathbf{v}_{p,x}(\mathbf{q})|) d^3\mathbf{q}%
\end{equation}
where $C_p(\mathbf{q})$ is the heat capacity, $v_{p,x}(\mathbf{q})$ is the group velocity along the $x$-direction, $\tau^{\mathrm{bulk}}_p(\mathbf{q})$ is the three-phonon scattering time, and $\Theta$ is the Heaviside function. We recall that the cumulative thermal conductivity is the thermal conductivity of phonons whose MFPs are below a given $\Lambda$~\cite{1995}. All the quantities appearing in ~\ref{cumulative} are taken from Ref.~\cite{esfarjani2011heat} and are not reported here for the sake of simplicity. The phonon dispersion curves and scattering times are obtained by means of harmonic and anharmonic force constants, which are extracted from DFT. The system's relaxation times are computed by using a uniform reciprocal space grid of $24\times 24\times 24$ points, harmonic force constants up to $5^{th}$ neighbors and cubic force constants up to first neighbors. We use the Local Density Approximation (LDA) from Perdew and Zunger~\cite{perdew1981self} with a energy cutoff of $40~\mathrm{Ryd}$. The MFP-BTE is solved for a set of $30$ MFPs, uniformly spaced on logarithmic scale from about $0.1~\mathrm{nm}$ to $100~\mathrm{\mu m}$. The MFP-BTE is discretized both in angle and space. The spatial discretization is achieved by means of the finite volume (FV) approach whereas the solid angle is discretized by means of the discrete ordinate method~\cite{mathur2002computation}. 

\begin{acknowledgement}

Research primarily supported as part of the Solid-State Solar-Thermal Energy Conversion Center (S3TEC), an Energy Frontier Research Center funded by the U.S. Department of Energy (DOE), Office of Science, Basic Energy Sciences (BES), under Award DE\-SC0001299. D.S. acknowledges funding from the MIT Deshpande Center for Technological Innovation.

\end{acknowledgement}



\providecommand{\latin}[1]{#1}
\providecommand*\mcitethebibliography{\thebibliography}
\csname @ifundefined\endcsname{endmcitethebibliography}
  {\let\endmcitethebibliography\endthebibliography}{}


\end{document}